\documentclass[preprint2]{aastex631}

\usepackage{multirow}
\usepackage{color} 

\received{June 11, 2022}
\revised{August 6, 2022}
\accepted{November 3, 2022}

\submitjournal{PASP}

\shorttitle{Unsupervised Galaxy Morphology}
\shortauthors{S.Wei et al.}
\graphicspath{{./}{figures/}}

\begin{document}

\title{Unsupervised Galaxy Morphological Visual Representation with Deep Contrastive Learning}

\correspondingauthor{Bo Liang}
\email{liangbo@astrolab.cn}

\author[0000-0002-3547-4025]{Shoulin Wei}
\affiliation{Faculty of Information Engineering and Automation, Kunming University of Science and Technology, Kunming, 650500, China}
\affiliation{Computer Technology Application Key Lab of Yunnan Province, Kunming University of Science and Technology, Kunming, 650500, China}

\author{Yadi Li}
\affiliation{Faculty of Information Engineering and Automation, Kunming University of Science and Technology, Kunming, 650500, China}

\author{Wei Lu}
\affiliation{Faculty of Information Engineering and Automation, Kunming University of Science and Technology, Kunming, 650500, China}

\author{Nan Li}
\affiliation{National Astronomical Observatories Chinese Academy of Sciences, Beijing, 100012, China}

\author{Bo Liang}
\affiliation{Faculty of Information Engineering and Automation, Kunming University of Science and Technology, Kunming, 650500, China}

\author{Wei Dai}
\affiliation{Faculty of Information Engineering and Automation, Kunming University of Science and Technology, Kunming, 650500, China}

\author{Zhijian Zhang}
\affiliation{Faculty of Science, Kunming University of Science and Technology, Kunming, 650500, China}



\begin{abstract}

Galaxy morphology reflects structural properties which contribute to understand the formation and evolution of galaxies. Deep convolutional networks have proven to be very successful in learning hidden features that allow for unprecedented performance on galaxy morphological classification. Such networks mostly follow the supervised learning paradigm which requires sufficient labelled data for training. However, it is an expensive and complicated process of labeling for million galaxies, particularly for the forthcoming survey projects. In this paper, we present an approach based on contrastive learning with aim for learning galaxy morphological visual representation using only unlabeled data. Considering the properties of low semantic information and contour dominated of galaxy image, the feature extraction layer of the proposed method incorporates vision transformers and convolutional network to provide rich semantic representation via the fusion of the multi-hierarchy features. We train and test our method on 3 classifications of datasets from Galaxy Zoo 2 and SDSS-DR17, and 4 classifications from Galaxy Zoo DECaLS. The testing accuracy achieves 94.7\%, 96.5\% and 89.9\% respectively. The experiment of cross validation demonstrates our model possesses transfer and generalization ability when applied to the new datasets. The code that reveals our proposed method and pretrained models are publicly available and can be easily adapted to new surveys\footnote{\url{https://github.com/kustcn/galaxy_contrastive}}.

\end{abstract}




\section{Introduction}
\label{sec:1}

Galaxy morphology provides the most intuitive features to inspect the structure of galaxies which has strong correlation with galactic star formation history. Massive early type galaxies tend to appear as elliptical. While spiral and irregular galaxies show evidence for on-going star formation. Galaxy morphological classification can lead to important astrophysical insights that helps to explore the formation and evolution of galaxies\citep{Wilman2012}. On the flip side, any theory of galaxy formation and evolution also have to explain the structures of dazzling galaxies.

The classification of galaxies dates back to the most famous Hubble sequence\citep{Hubble1926}. De Vaucouleurs system\citep{Vaucouleurs1959} extends Haber sequence to form 4 main categories: Elliptical, Lenticular, Spiral and Irregular. In particular, Spiral type be subdivided into Bars, Rings and Spiral Arm which can still be divided into subtler classification according to the tightness of spiral arms. 

The traditional methods of morphological classifications based on visual inspection and fitting light profiles have achieved remarkable success. However, these methods are infeasible in the era of large data volume. The forthcomming survey projects will reach increasing depths and widths over nearly the entire extragalactic sky and produce massive galaxies. The Large Synoptic Survey Telescope (LSST) will produce billions of galaxies covering the entire southern sky in multiple bands \citep{lsst2019}. The space-borne telescopes, e.g., Euclid \citep{racca2016} and the Chinese Space Station Telescope (CSST), \citep{zhan2021} will cover 15000 and 17500 square degrees of the celestial sky respectively and is expected to detect billions of galaxies.

In response to the coming exponential growth data volume of galaxies, automated morphological classification has become an inevitable approach. In the past few years, we have witnessed the tremendous success of the Convolutional Neural Network (CNN) based Deep Learning (DL) in various computer vision tasks. Meanwhile, CNNs also have been extensively studied for morphological classification of galaxies. In a Kaggle competition of galaxy morphological classification on Galaxy Zoo 2 \citep{willett2013}, CNN based model finished in first place out of 326 participants \citep{Dieleman2015}. \citet{Cheng2022} carry out a comparison between CNN and the traditional machine learning methods including K-nearest neighbour, logistic regression, Support Vector Machine, Random Forest and Neural Networks over the Dark Energy Survey (DES) data and Galaxy Zoo 1. Their results show that CNN has the best performance. \citet{Khalifa2018} proposed a Deep Galaxy V2 based on Deep Convolutional Neural Networks for galaxy classification. The architecture was trained over 4238 images taken from the EFIGI catalogue \citep{Baillard2011} and achieved a 97.772\% testing accuracy on 3 classifications (Elliptical, Spiral, and Irregular). \citet{Zhu2019} proposed a variant of residual network, Resnet-26, which classifies 28790 galaxy image samples from Galaxy Zoo 2 into 5 categories. The results show that Resnet-26 achieves the best performance with 95.2083\% accuracy, compared with AlexNet, VGG, Inception and Resnet. \citet{Katebi2019} used Capsule Network (CapsNet) for regression and predicted probabilities for all of the questions in the Galaxy Zoo project and trained a CapsNet classifier that outperforms the baseline CNN by 36.5\% error reduction. \citet{Gupta2022} introduced a continuous depth version of the Residual Network called Neural Ordinary Differential Equations (NODE) which obtained an accuracy between 91\%–95\% depending on the classifications. DL methods used for galaxy morphological classifications are discussed in  \citet{tuccillo2016deep}, \citet{khan2019deep}, \citet{Ghosh2020}, \citet{bhambra2022}, \citet{zzr2022RAA} and \citet{Vavilova2022}.

The machine learning methods in galaxy morphological classification are still dominated by supervised learning at present. However, acquisition of large training sets with labelled data is one of the key challenges, especially when approaching a new survey project. Transfer learning, to some extent, allows us to deal with a new domain without labelled data by leveraging the already existing trained model with labelled data of some related task. \citet{Dominguez2018} proposed a transfer learning method with a CNN model trained with Sloan Digital Sky Survey (SDSS) data. When applying the models directly to unlabeled Dark Energy Survey (DES) data, the accuracy reaches approximately 90\%. In \citet{Variawa2022}, the proposed method used the pre-trained ResNet50 based model fined-tuned on the Galaxy Zoo 2 and EFIGI with expert labelled data. Using 9 Hubble classes, this model achieved an accuracy of 30.50\% on the Revised Shapley-Ames (RSA) catalogue as a test data. Although these results have proved the effectiveness of transfer learning from one survey to another, the models still need to be fine-tuned to generalize well to the data set with similar distribution but would fail catastrophically on unexpected or atypical inputs like totally different redshift distribution. 

Compared with supervised learning, unsupervised learning that aims to uncover the internal structure of a data set without the need for any label will be the solution for automated morphological classification of the new and forthcoming surveys. \citet{yangh2022RAA} adopted Convolutional Auto-Encoder (CAE) and attention mechanism to detect astronomical outliers in the data of galaxy images in an unsupervised manner. \citet{Schutter2015} used unsupervised analysis to quantitatively obtain similarities between the different morphological types using merely the galaxy images, whereas the purpose of the method is not to automatically classify galaxies, and the method rely heavily on the features extraction and analysis. \citet{Martin2020} constructed an unsupervised model based on patches for automatic segmentation and labeling of galaxy images where Growing Neural Gas (GNG) is used for features extraction, Hierarchical Clustering (HC) for a hierarchical representation and Connected Component Labeling for construction of object feature vectors. Galaxy morphological classification is obtained by comparing the similarity between feature vectors. \citet{Cheng2020} combined feature extraction with vector-quantized variational autoencoder (VQ-VAE) and HC for unsupervised machine learning to explore galaxy morphological analysis. The test accuracy reached 87\% when perform a binary classification over two large preliminary clusters.

Inspired by the success of contrastive learning in computer vision \citep{he2020momentum, hjelm2018learning, van2018}, we are interested in exploring whether it could also be used for unsupervised galaxy morphological visual representation. \citet{Hayat2021} have applied CL to perform galaxy morphological classification and photometric redshift estimation on multiband galaxy photometry from SDSS. Their results achieve the accuracy of supervised models while using
2-4 times fewer labels for training. \citet{Sarmiento2022} adopted Simple framework for Contrastive Learning of visual Representations (SimCLR) to galaxy morphological classification on SDSS galaxy images with $z < 0.15$. The accuracy achieved 85\%, which was similar to the results of their supervised learning approach.

Based on contrastive learning, we propose an unsupervised learning network which includes a well-designed encoder to learn multi-hierarchy feature representation. We carried out training on the data sets of Galaxy Zoo 2, SDSS-DR17 and Galaxy Zoo DECaLS, with test accuracy of classification reaching 94.7\%, 96.5\% and 89.9\% respectively. In order to verify the robustness of the model, we performed cross validation in that way the model was trained with one of the three datasets, and tested on the other two datasets. The accuracy of cross validation have a less reduction and still achieved approximately 90\%. The results show that the model possesses high transfer and generalization ability, and could be applied to the new surveys. The results of experiments performed with hierarchical clustering demonstrates that visual representation learned include detailed features on which clustering can identify subtler classifications.

The remainder of this paper is organized as follows: Section \ref{sec:dataset} describes morphological catalogue used in this work and data preprocessing. Section \ref{sec:methods} describes the architecture of the proposed network and our methodology. Section \ref{sec:experiments} reports the experiments performed and discusses the results. Finally, Section \ref{sec:conclusions} concludes the paper and describes ideas for future work.

\section{DATASET }\label{sec:dataset}
In this paper, we train the proposed model and test the performance with sample images from GZ2, GZ DECaLS and SDSS-DR17. In this section we describe the galaxy catalogues, the sampling strategy and threshold configurations of the constructed datasets used for this study.

\subsection{Galaxy Zoo}

Galaxy Zoo (GZ) is an online crowdsourcing project which invites volunteers to visually inspect and classify these galaxies via the internet (\url{zooniverse.org}) \citep{Lintott2008}. Since 2007, volunteers involved have performed billion classification tasks. Their work has greatly promote the development of galaxy morphological classification, and led to a robust database of galaxy classifications \citep{Lintott2011} including GZ2, GZ Hubble, GZ CANDELS and et al. At present, GZ has gone to the 4th phase launched in 2012 \citep{Simmons2016}.

\begin{figure*}
	\includegraphics[width=1\linewidth]{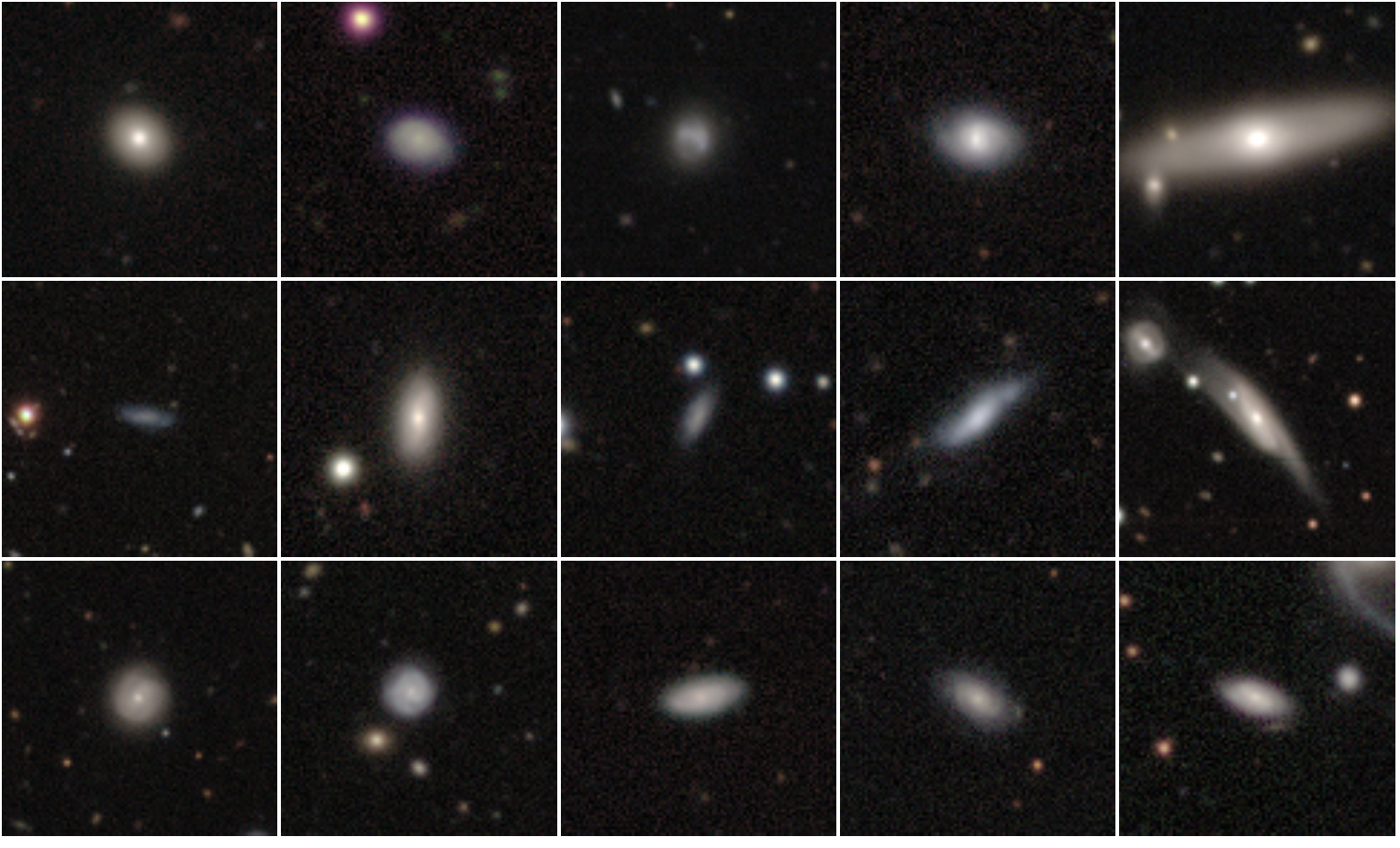}
    \caption{Examples of a sample of galaxies from GZ2. The first row shows five images labeled as Round type. but it can be seen that the first two have little difference in contour information with elliptical galaxies, the third and fourth have slightly spiral features, and the last one is more similar to the lateral and cigar shape, but these five images are all classified as round galaxies. Similarly, the labels in the second row are lateral and cigar shaped. The first three are similar to elliptical galaxies, while the last two have spiral arms and are more suitable for spiral galaxies. The third row is labeled spiral galaxies, but because the spiral arms are less distinctive, the first two are more circular and the last three more elliptical.}
    \label{fig:fig_gz}
\end{figure*}

In this work, the two datasets used are drawn from GZ2 and Dark Energy Camera Legacy Survey (DECaLS) in GZ catalogues. The GZ2 catalogue presents morphological classification of more than 300,000 nearby galaxies from SDSS \citep{willett2013}. The GZ DECaLS catalogue taken from the Victor M. BLanco 4M telescop have published classifications of 314,000 galaxies so far \citep{walmsley2022galaxy}. The GZ classifications are based on 
the volunteers' answers to questions of binary decision trees (The details of decision trees are depicted at \url{https://data.galaxyzoo.org/gz_trees/gz_trees.html}). Volunteers are asked various questions, such as 'How rounded is it?' and 'Could this be a disk viewd edge-on?'. Various volunteers' answers for one image produce a vote fraction for each feature. For example, a image with $F_{edge-on} = 0.7$ represents that 70\% volunteers answered 'Yes' to the question 'Could this be a disk viewd edge-on?' of this image. Unsupervised learning works on unlabeled data when training models. Without this powerful correction, it is necessary to enhance the differences between different classes of datasets. Selection with moderate thresholds such as $F_{edge-on} > 0.5$, $F_{features,disk} > 0.5$ and $F_{not-clumpy} > 0.5$ may lead to less differences between classes (as shown in Fig.\ref{fig:fig_gz}). Therefore, we prune these datasets and merely select those images, which are classified with high vote fractions in their respective classes. The detailed of vote fraction threshold for selection is listed as Table.\ref{tab:data_table}. After pruning, we have a total of 7168 images of GZ2 and 16284 images of GZ DECaLS. The samples of GZ2 are in the classifications of Elliptical, Edge-on and Spiral. Each image of GZ DECaLS is divided into four classes (Round, Elliptical, Edge-on, and Spiral). 

\subsection{SDSS}
This work uses the third dataset acquired from the latest 17th public data release of SDSS Phase IV (SDSS-IV) \citep{dominguez2022sdss}. The Catalog Archive Server (CAS) of the SDSS provides a Structured Query Language (SQL) web-based interface for researchers to retrieve galaxies of interest. The full SQL we used to retain about each sample, can be found at \url{https://github.com/kustcn/galaxy_contrastive/blob/master/prepdata/sdss_pdr17_query.sql}, and can be run through the SDSS data access site at \url{http://skyserver.sdss.org/dr17/SearchTools/sql} (accessed Jan 7, 2022). 

The detailed selection criterias for SDSS are described in Table.\ref{tab:data_table}. \emph{zns.p\_el} represents the probability of the sample considered as elliptical, \emph{zns.p\_cw} represents the probability of having a clockwise spiral shape, and \emph{zns.p\_acw} represents the probability of having a counterclockwise spiral. In total, this sample from SDSS includes 9914 galaxies classified into the same three classes as the samples from GZ2. In addtion, the global conditions for all SDSS samples include \emph{g.clean}$>$1, \emph{g.scale}=0.2, and
\emph{zns.nvote}$>$20 that indicates selecting the objects with the greater degree of clean than 1, the 0.2 scale factor of each image and the greater number of classified votes of each image than 20, respectively.

\begin{table*}
	\centering
	\caption{The vote fraction threshold and selection criterias of samples from GZ2, GZ DECaLS and SDSS.}
	\label{tab:data_table}
	\begin{tabular}{lcccccc} 
		\hline
		\multirow{2}{*}{Type} & \multicolumn{2}{c}{GZ2} & \multicolumn{2}{c}{SDSS} & \multicolumn{2}{c}{GZ DECaLS}\\
		& Selection & Count & Selection & Count & Selection & Count\\
		\hline
		 \multirow{2}{*}{Round}
		 &  &  &  &  & $F_{smooth} \ge 0.9$ & \\
		 &  &  &  &  & $F_{completely round} \ge 0.9$ & 2926\\
		\hline
		 \multirow{2}{*}{Elliptical}
		 &  $F_{smooth} \ge 0.7$ &  &  &  & $F_{smooth} \ge 0.95$ & \\
		 &  $F_{in-between} \ge 0.7$ & 2203 & \emph{zns.p\_el} $ \ge 0.8$ & 3000 & $F_{in-between} \ge 0.95$ & 5263\\
		\hline
		 \multirow{2}{*}{Edge}
		 & $F_{features,disk} \ge 0.7 $ &  &  &  &$F_{features,disk} \ge 0.79 $ & \\
		 & $F_{edge-on/yes} \ge 0.7 $ & 2058 & \emph{zns.p\_edge} $ \ge 0.8 $ & 2994 & $F_{edge-on/yes} \ge 0.79 $ & 5871 \\
		\hline
		 \multirow{3}{*}{Spiral} & $F_{features,disk} \ge 0.9$ &  & &  & $F_{features,disk} \ge 0.9$ & \\
		 & $F_{edge-on/no} \ge 0.9$&   & \emph{zns.p\_cw} $ \ge 0.8$  & 1919 & $F_{edge-on/no} \ge 0.9$ & \\
		 & $F_{spirals/yes} \ge 0.9$ & 2907 & \emph{zns.p}\_acw $ \ge 0.8$  & 2001 & $F_{spirals/yes} \ge 0.9$ & 2224\\
		\hline
		Count  &  & 7168 &  & 9914 &  & 16284\\
		\hline
	\end{tabular}
\end{table*}

\subsection{Pre-processing}

\begin{figure}
	\includegraphics[width=\columnwidth]{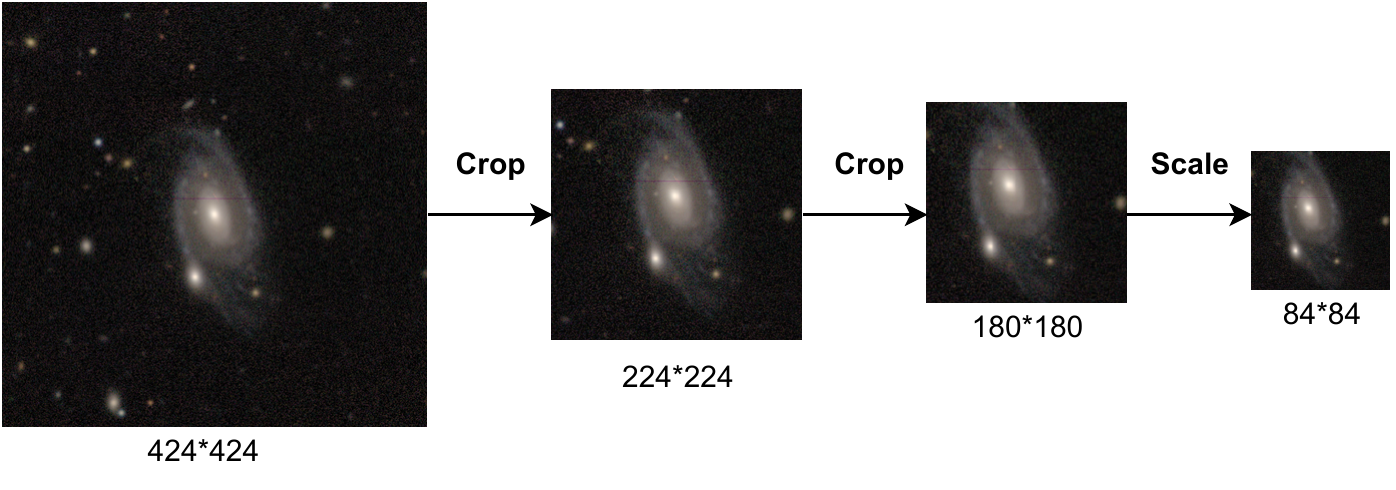}
    \caption{The steps of pre-processing procedure for GZ2 and GZ DECaLS. The first crop acts on the centre of image. In order to increase randomness and disturbance, the $180\times180$ image is randomly cropped out of the $224\times224$ image. The scaling peforms on GZ2 and GZ DECaLS from $180\times180$ to $84\times84$.}
    \label{fig:fig_preprocessing}
\end{figure}

Each image from GZ2 and GZ DECaLS is an RGB composite image with $424\times424$ pixel in size. The galaxy of interest is generally located at the centre of the image. The image quality in GZ2 is relatively poor, with an average size of a dozen KB, while the average size of a single image in GZ DECaLS reaches 300KB. This means that the resized pixels with small scale in GZ2 result in seriously distorted images. All images are feed to the training model as $84\times84$ by 3-channel RGB values, that is a reasonable compromise between image quality and computation efficiency. In practice, we centrally crop and scale the image to that size. Fig.\ref{fig:fig_preprocessing} shows the steps of pre-processing procedure for GZ2 and GZ DECaLS, which allows the main information to be contained in the centre of the image and eliminates all random noises like some other secondary object. In the selection of SDSS, we uniformly define scale=0.2 and cutout size with $128\times128$, then the original downloaded images have the size of $128\times128$. Therefore, the pre-processing procedure for SDSS only includes scaling from $128\times128$ to $84\times84$.

It notes that unsupervised learning does not require labels in the training phase, we only use the labels in the testing phase to verify the accuracy of our proposed method.

\section{METHODOLOGY}\label{sec:methods}
\begin{figure*}
	\includegraphics[width=\linewidth]{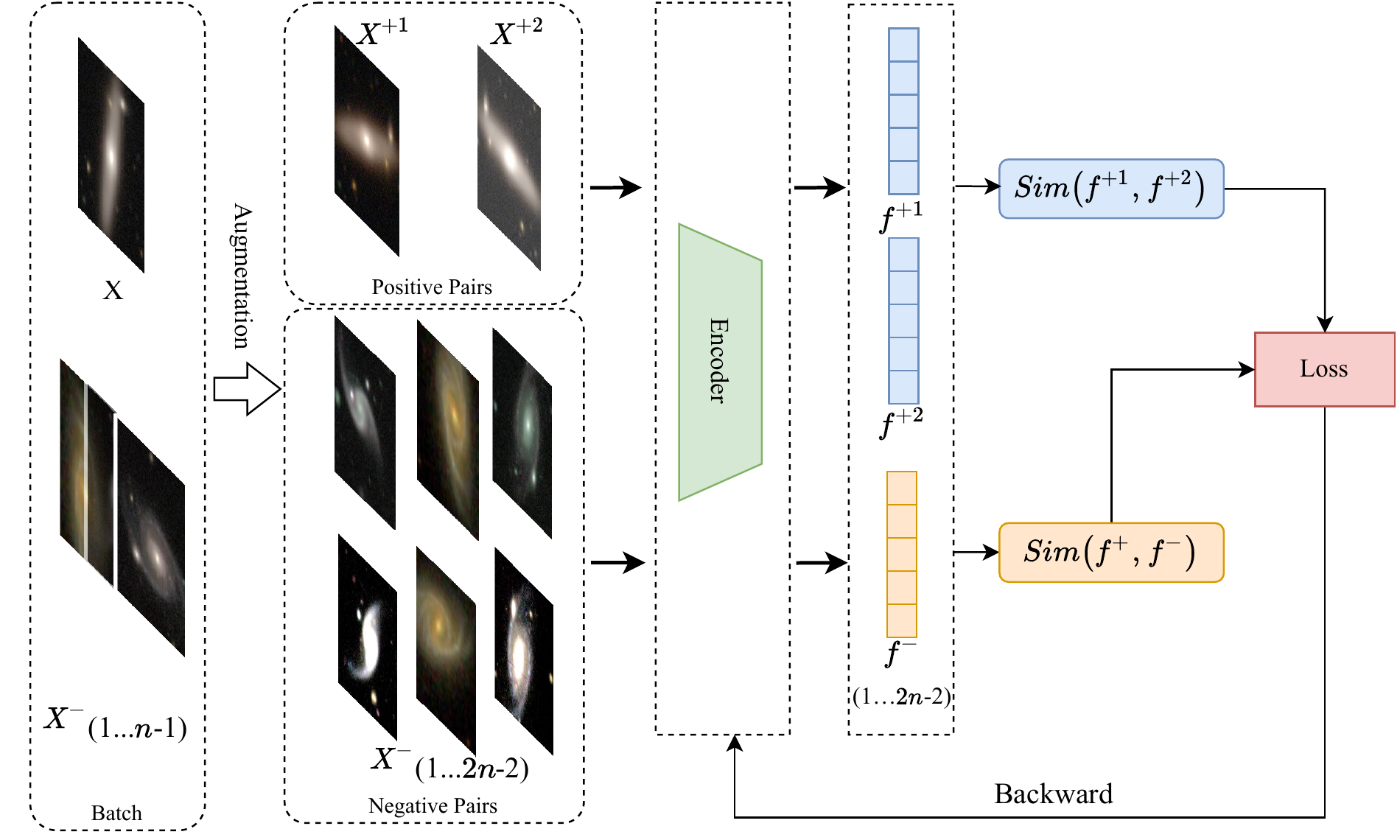}
    \caption{The overall architecture of the proposed network, which learns representation with multi-hierarchy features by \emph{Encoder} and backward the loss calculated from the similarities between the positive pair and negative pairs.}
    \label{fig:fig_framework}
\end{figure*}

In this section, we first briefly introduce the overall architecture of our proposed method as shown in Fig.\ref{fig:fig_framework}. Then, we present the strategies of data augmentation. Next, we discuss the construction of encoder that converges the multi-hierarchy features with deep learning. Finally, we describe the definition of loss function.
\subsection{Architecture}

The architecture of our method is mainly based on a strong  framework for contrastive learning SimCLR \citep{chen2020}. First of all, for each image in a batch, we apply two random data augmentation to obtain two different views. For one image $X$, the augmented views ($X^{+1}$ and $X^{+2}$) are called the positive pair of this image. While the views of other images are called corresponding negative pairs. Then, \emph{Encoder} learns feature representation that best describes each view and produces 1-dimensional vector with size 128 for each view. Accordingly, $f^{+1}$ and $f^{+2}$ is representation for $X^{+1}$ and $X^{+2}$, $f^{-}$ for negative pairs. Finally, the similarity \emph{Sim} between feature representations is measured with cosine similarity. The goal of contrastive learning is to increase the similarity between instances from the positive pairs and decrease the similarity between the positive and negative pairs. Instance-wise contrastive learning achieves this objective by optimizing a contrastive loss function (discussed in Section \ref{sec:loss}) and back-propagation through \emph{Encoder} to update the network parameters.
\subsection{Data Augmentation}
\begin{figure}
	\includegraphics[width=\columnwidth]{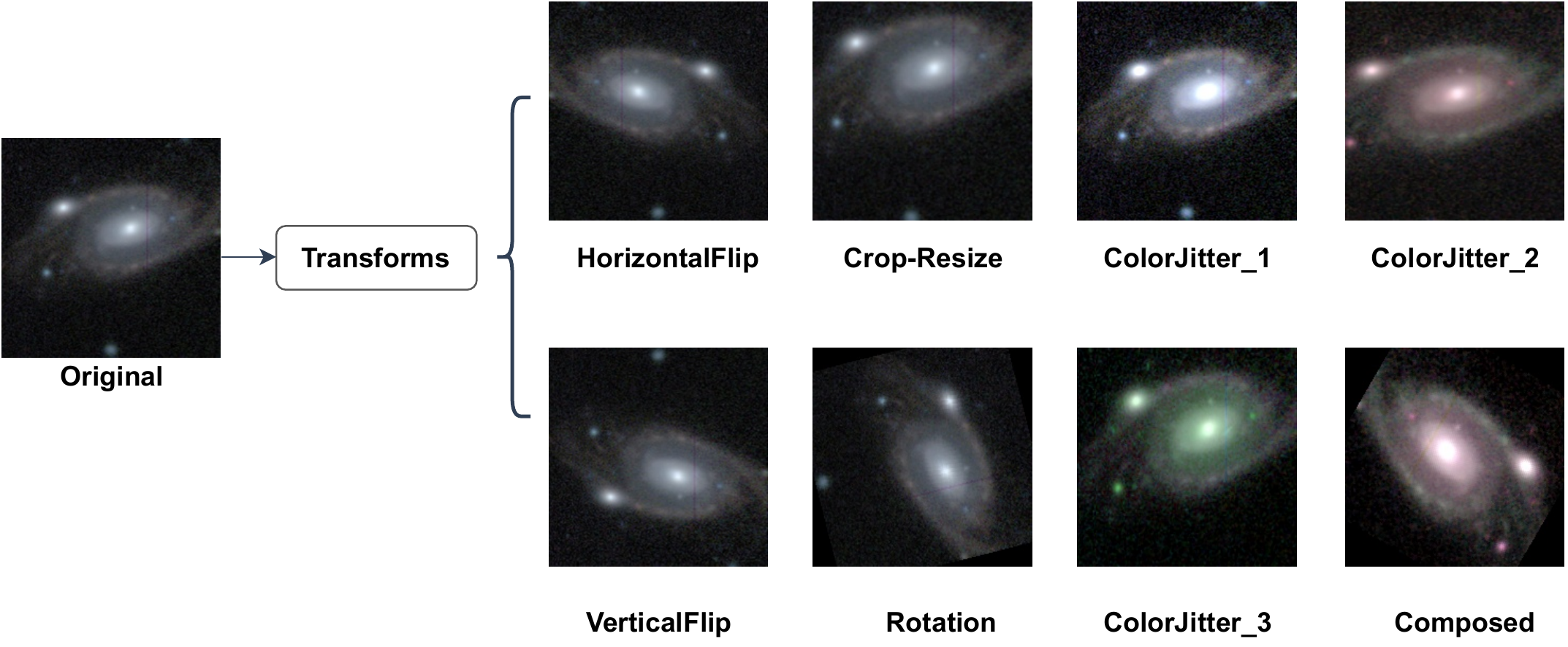}
    \caption{The original image and the eight sample results after applying data augmentation used in this work. }
    \label{fig:fig_aug}
\end{figure}

In computer vision tasks of supervised learning, data augmentation generally is used to expand the size of a training set and increase variability by creating modified data from the existing data. While data augmentation has played core rule of creating pretext tasks for training a contrastive learning network. In this pretext, we construct the positive pair and negative pairs by randomly augmenting twice for all the image in a batch. Fig.\ref{fig:fig_aug} shows the original image and the images after applying these augmentation:

\begin{enumerate}
    \item HorizontalFlip and VerticalFlip: a image is flipped with a probability of 0.5 horizontally and vertically.
    \item Crop and Resize: random crop with a scale factor of 0.8, and then scale to the original size ($84\times84$).
    \item ColorJitter: random adjustment of brightness, contrast, saturation and hue of the image. The corresponding values are 0.2, 0.5, 0.5 and 0.4. Taking 0.2 as an example, it means to randomly change the brightness to 80\%(1-0.2) - 120\%(1+0.2) of the original image brightness.
    \item Rotation: random rotation with an angle sampled uniformly between $0^{\circ}$ and $360^{\circ}$.
    \item Composed: composition with a series of transformations mentioned above.
\end{enumerate}

\subsection{Encoder}
\label{sec:encoder}
\begin{figure*}
	\includegraphics[width=1\linewidth]{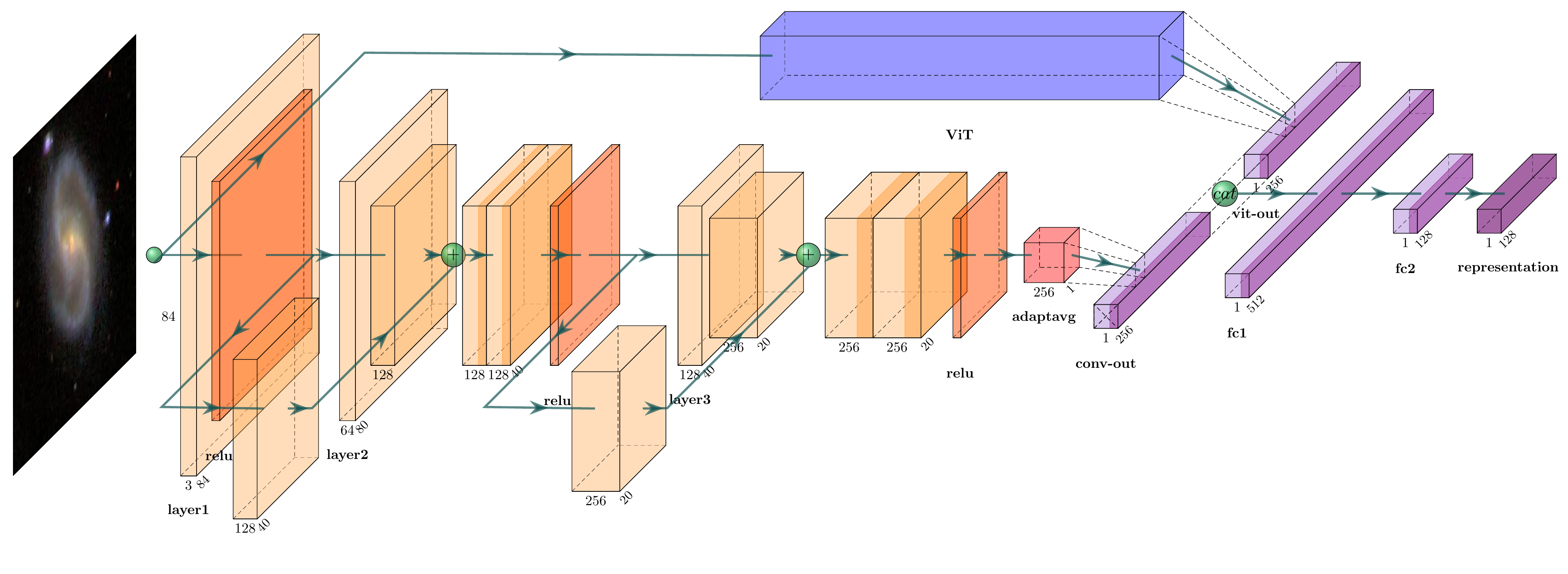}
    \caption{The illustration of components in \emph{Encoder}, which use ResNet-18 as the backbone network fusing with ViT to learn multi-hierarchy features.}
    \label{fig:fig_encoder}
\end{figure*}

The goal of \emph{Encoder} is to identify latent structure by using pixel data alone. Especially, galaxy images are relatively lacking in semantic information, with monotonous colors and single contour features. Some galaxies with strong luminosity even have no obvious contours, but only scattered halos. Therefore, the feature extraction of galaxy images is more susceptible to noise interference than common datasets of computer vision task. We choose the current start-of-the-art transformers \citep{Vaswani2017} combined with CNN for learning multi-hierarchy features representation of the input images. Instead of using only coarse semantic features as conventional CNNs, we resort to capability of Vision Transformer (ViT) on global spatial information and CNN on local information for capturing rich representation of galaxies. As shown in Fig.\ref{fig:fig_encoder}, we use ResNet-18 \citep{he2016deep} as the backbone network. As discussed in \citet{Raghu2021}, ViT incorporates more global information than ResNet at lower layers. The training data is fed into the ResNet-18 based and the ViT sub-networks in parallel. Thus, multi-hierarchy features depends on fusing the learned features from these two sub-networks. 

Considering that galactic images have less semantic information and the classification mainly relies on contour features, we reduce the depth of the network model (remove the forth layer of ResNet-18), in order to quickly extract contour features. We also remove the maximum pooling in first convolutional layer. Since maximum pooling may filter out slightly darker structure information of galaxies with bright core structures, and increase the interference of galaxies with higher luminosity. The rectified linear units (RELU), \citep{agarap2018deep} activation is used after each convolution layer except the last one. 

The output size of the ResNet-18 based sub-network after a adapted-average pooling is with size of $1\times256$. The output size of ViT is with same size of $1\times256$. Before fed into a Multilayer Perceptron (MLP) block, the fusing of features with size of $1\times512$ forms by concatenating features from these two sub-networks. The MLP block with two fully-connection layers is used to map the features extracted by the backbone to the latent space with size of $128\times1$. Instead of a linear layer, a MLP block will retain detailed information related to data augmentation while the role of a linear layer is to remove this information \citep{chen2020}. We include the layers details of \emph{Encoder} in Table.\ref{tab:layers}. 

\begin{table*}
	\centering
	\caption{The layers details of \emph{Decoder}, which consists of three convolution blocks, a visual transformers block, and two fully connected layers. The input shape, output shape and the number of parameters in each layer are also shown.}
	\label{tab:layers}
	\begin{tabular}{llccc} 
		\hline
		Block & Layer Type & Input shape  &  Output Shape & Parameters\\
		\hline
		{layer1} & Covn2D(7*7) & (3, 84, 84) & (64, 80, 80) & 9408 \\
		\hline
		\multirow{5}{*}{layer2} & Covn2D(3*3) & (64, 80, 80) & (128,40,40) & 73728 \\
		& Covn2D(3*3) & (128,40,40) & (128,40,40) & 147456\\
		& Covn2D(1*1) & (64, 80, 80) & (128,40,40) & 8192 \\
	    & Covn2D(3*3) & (128,40,40) & (128,40,40) & 147456\\
	    & Covn2D(3*3) & (128,40,40) & (128,40,40) & 147456\\
		\hline
		\multirow{5}{*}{layer3} & Covn2D(3*3) & (128,40,40) & (256, 20, 20) & 294912 \\
		& Covn2D(3*3) & (256, 20, 20) & (256, 20, 20) & 589824\\
		& Covn2D(1*1) & (128,40,40) & 256, 20, 20) & 32768 \\
		& Covn2D(3*3) & (256, 20, 20) & (256, 20, 20) & 589824\\
		& Covn2D(3*3) & (256, 20, 20) & (256, 20, 20) & 589824\\
		\hline
		& Avg pooling & (256, 20, 20) & (256, 1, 1) & 0 \\
		\hline
		& Flatten() & (256, 1, 1) & (256) & 0 \\
		\hline
		\multirow{4}{*}{VIT} 
		& Rearrange & (3, 84, 84) & (7*7,432) & 0 \\
		& Linear(embed) & (49+1, 432) & (50,128) & 16384\\
		& 6*Attention & (50,128) & (50,128) & 1179648\\
		& Linear(mlphead) & (128) & (256) & 16384 \\
		\hline		
		\multirow{2}{*}{Projector} & Linear & (512) & (512) & 262144 \\
		& Linear & (512) & (128) & 65536\\
		\hline
		& Total: &  &  & 4187328 \\
		\hline
	\end{tabular}
\end{table*}
\subsection{Loss Function}
\label{sec:loss}
InfoNCE \citep{van2018} loss was used as a contrastive loss, which aims to pull positive examples to be closer and push negative examples to be farther apart. We take Cosine similarity as a distance metric between the representation vectors. The similarity function is defined as equation~(\ref{eq:sim}).
\begin{equation}
Sim(f_i,f_j) = \frac{f^T_i f_j}{\Vert f_i \Vert_2 \Vert f_j \Vert_2}
\label{eq:sim}
\end{equation} 
$\Vert f_i \Vert_2$ and $\Vert f_j \Vert_2$ is the L2 regularization of the representation vector $f_i$ and $f_j$. \citep{Wang2020} argued that L2 regularization of the representation vector can improve the training model. For the $i$-th instance, the loss function is defined as equation~(\ref{eq:loss}).
\begin{equation}
\mathcal{L}_{i} = - \log \frac{\exp \left( Sim(f^{+1}_i,f^{+2}_i) / \tau\right)}{\sum_{j=0}^{K} \exp \left(Sim(f_i,f_j) / \tau\right)}
\label{eq:loss}
\end{equation} 
Where, $f^{+1}_i,f^{+2}_i$ represent the corresponding representation vectors of positive pair. Take a batch of $n$ images, the number of negative samples $K$ is $2n-2$. $\tau$ called temperature is a hyper-parameter which plays a role in controlling the strength of penalties on negative samples \citep{Wang_2021_CVPR}. The final loss is an arithmetic mean of the loss for each sample in the batch. It can be defined as equation~(\ref{eq:loss2}).
\begin{equation}
\mathcal{L} = \frac{1}{N} \sum_{k=1}^{N} \mathcal{L}_{k}
\label{eq:loss2}
\end{equation}
The objective of model learning is to minimize the loss function. It can be seen that the molecular part of equation~(\ref{eq:loss}) encourages the higher similarity between the positive samples. In the denominator part, the similarity between positive and negative samples is encouraged to be as low as possible. In this way, during the optimization process, the model can be trained by the loss function guidance to achieve the desired goal.
\section{Experiments}\label{sec:experiments}
\subsection{Training}
We use Stochastic Gradient Descent (SGD) as our optimizer. The size of a batch is set to 256, a momentum is 0.9, and the maximum number of epochs is 50. The temperature $\tau$ is 0.1, and weight decay is 0.0001. We warm-up the network in the first 20 epochs by only using the InfoNCE loss. In virtue of Learning Rate Finder \citep{smith2017cyclical}, the initial learning rate is set to 0.01, and reduced with a cosine curve during the 50 epochs. We implemented our model on Pytorch. We train the model on the three datasets of GZ2, GZ DECaLS and SDSS-DR17 respectively. Each of dataset is splited into three different sets, a training, a validation, and a test set, which are randomly drawn from the total sample of galaxies. The training set consists of 80\%, the validation set of 10\%, and the test set of 10\% of the data. The model takes about 1 to 2.5 hours for training  50 epochs on a NVIDIA TITAN V GPU. All of the experiments were performed on a CentOS 7.7 server configured with dual Intel Xeon CPU E5-2660 v4, 3.4-GHz maximum frequency, and 512 GB of RAM.
\subsection{Evaluation Metric}
To evaluate the classification performance of the proposed model, we adopt k-nearest neighbors algorithm($k$NN) provided by faiss \citep{johnson2019billion} for efficient classification, agglomerative hierarchical clustering by scikit-learn \citep{pedregosa2011scikit} for clustering. $k$NN performs a voting mechanism to determine the class of a data point. For a unseen data point, $k$NN computes the nearest $k$ neighbors of it and performs a voting mechanism to determine the class . Among the $k$ neighbors, one of the classes accounts for the highest proportion, then this class is regarded as the predicted label of the data point. Accuracy is defined as the percentage of the number of correct predicted classifications compared to the true labels from the total samples. 

NMI \citep{Danon2005} and ARI \citep{Santos2009} is introduced to evaluate the performance of clustering results. To some extent, NMI and ARI reflects the level of information correlation between clustering results and real labels. We use the implementations of NMI and ARI via the functions \emph{normalized\_mutual\_info\_score} and \emph{adjusted\_rand\_score} in scikit-learn \citep{pedregosa2011scikit}. These two functions measure the similarity of the two assignments (predicted labels and real labels). The value range of NMI is $[0,1]$, and the value range of ARI is $[-1,1]$. The closer the value is to 1, the closer the clustering results are to the real labels.

\subsection{Performance}
\begin{figure*}
	\includegraphics[width=\linewidth]{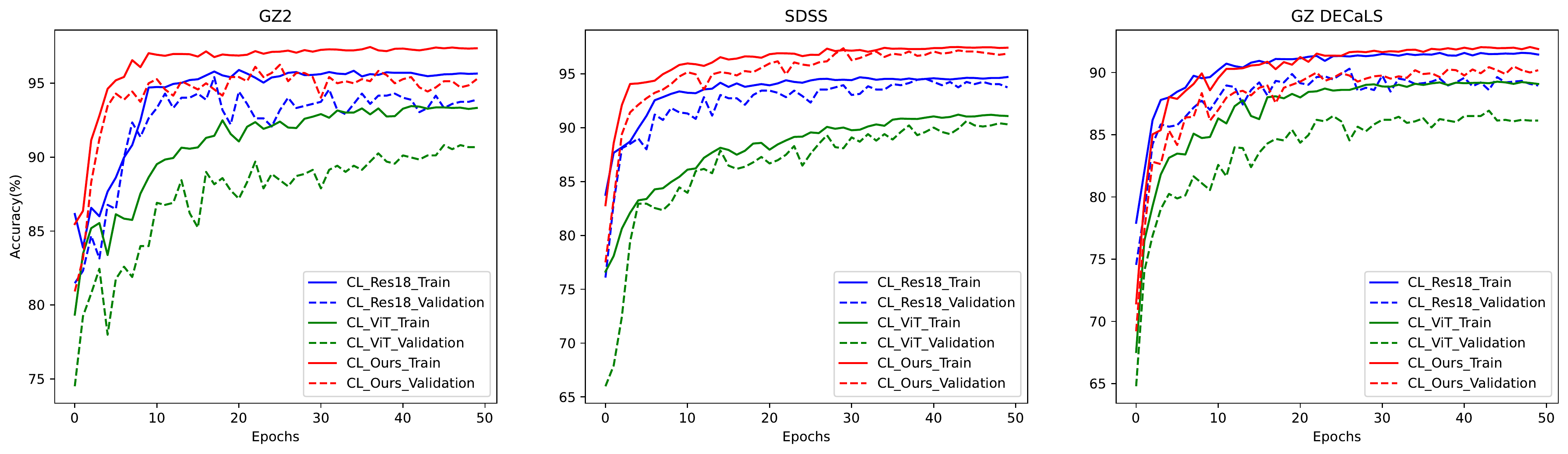}
    \caption{Performance comparison of the accuracy of training and validation along with the epochs. CL\_Res18, CL\_ViT and CL\_Ours represents the CL methods using ResNet-18, ViT and our proposed model as encoder.}
    \label{fig:fig_train_val_acc}
\end{figure*}

We performed the comparative performance tests of using ResNet-18, ViT and our proposed model as encoder. The tests were conducted through 50 epochs on the three datasets. The results are shown in Fig.\ref{fig:fig_train_val_acc}. Compared with only using ResNet-18 and ViT, our improved model achieves about 0.44\%-2.73\% increase in training accuracy and about 1.22\%-3.13\% increase in validation accuracy. 

\begin{figure*}
	\includegraphics[width=\linewidth]{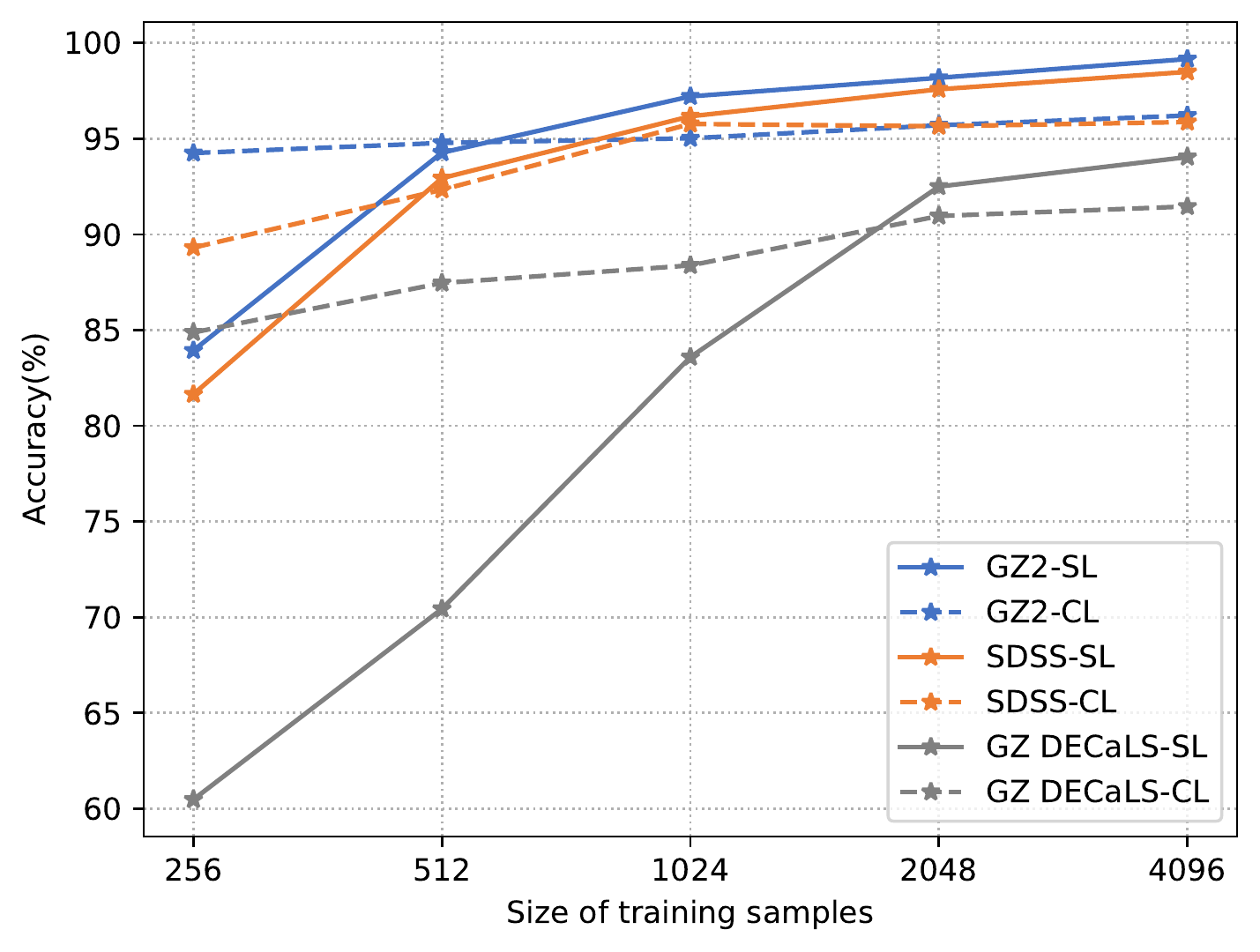}
    \caption{Performance comparison between our CL-based method and the SL approach. The panels demonstrate the testing accuracy along with the different size of training samples from GZ2, SDSS and GZ DECaLS respectively.}
    \label{fig:fig_with_sl}
\end{figure*}

To explore the performance comparison with supervised learning (SL), we performed the tests on the different size of training samples extracted from the three datasets. For the training subsets of GZ2 and SDSS, the samples are uniformly distributed in three categories (Elliptical, Edge-on and Spiral). For GZ DECaLS, the subsets are uniformly sampled from four categories(Round, Elliptical, Edge-on, and Spiral). To facilitate a fair comparison, SL approach is trained from scratch with the same architecture of our encoder. As shown in Fig.\ref{fig:fig_with_sl}, when the size of training subset is less than 512, our CL-based method achieves better performance than SL. Especially for GZ DECaLS, trained on 256 samples, our CL-based method can achieve an accuracy of about 85\%, while the accuracy of SL is only 60\%. Moreover, when the size of training samples increases to nearly 2000, our CL-based method is still ahead of SL. For the subsets from GZ DECaLS include four classifications, that means SL requires more samples to learn sufficient features. When sufficient samples feed, the performance of our CL-based method is with a few percentage points short to SL.

\begin{figure*}
	\includegraphics[width=1\linewidth]{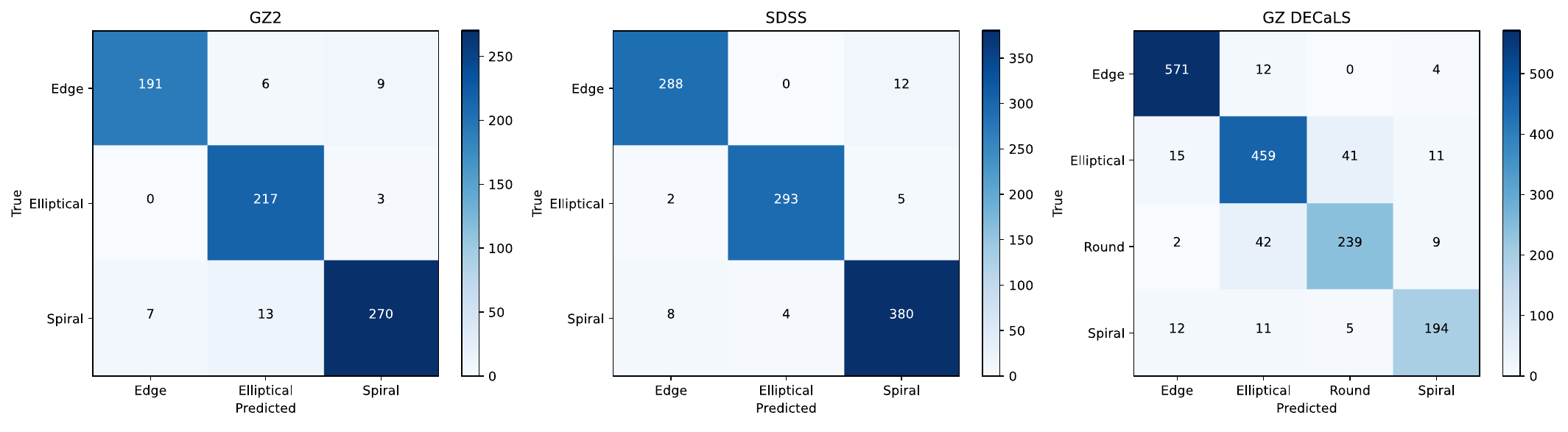}
    \caption{The confusion matrixes of testing results on GZ2, DECaLS and SDSS-DR17, where the x-axis corresponds to the predicted label from the network and the y-axis to the visually assigned label.}
    \label{fig:fig_confusion_matrix}
\end{figure*}

\begin{figure*}
	\includegraphics[width=1\linewidth]{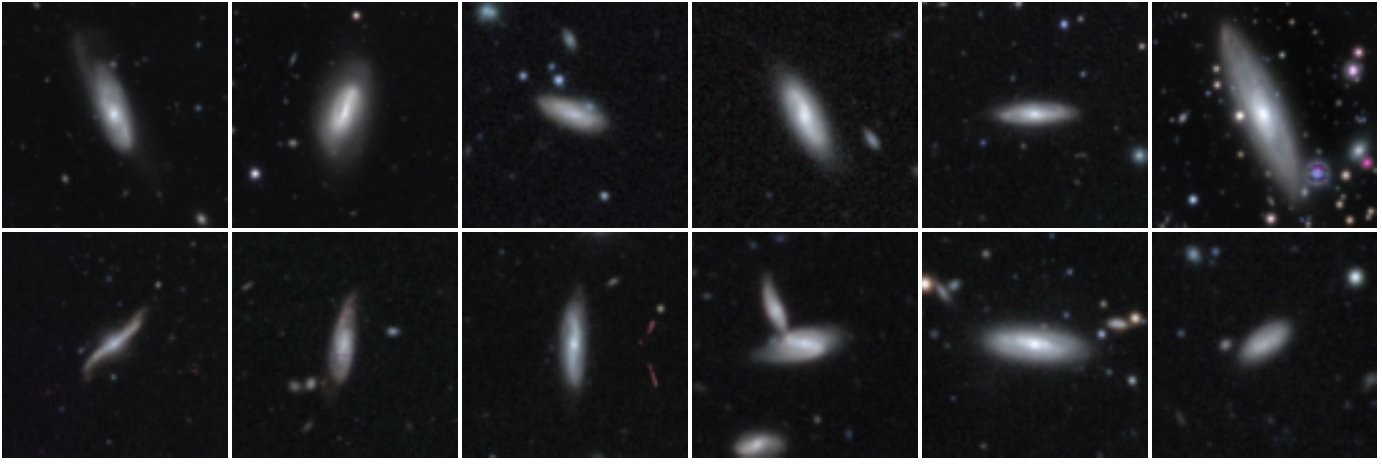}
    \caption{The samples mistakenly predicted as Edge-on galaxies from the test set of DECaLS, which are originally voted as spiral galaxies.}
    \label{fig:fig_error_labeled}
\end{figure*}

We also perform $k$NN under $k=1,5,10,15,20$ directly on the learned representations from our trained model. The accuracies on testing datasets are describe in Table \ref{tab:acc_table}. Note that the testing accuracy is the median accuracy calculated from 5 different runs. In order to analyze the specific classification of the test set in the three data sets in detail, we illustrate the confusion matrixes upon the three test sets in Fig.\ref{fig:fig_confusion_matrix}. Taking DECaLS as a example, there are 12 spiral galaxies that are mistakenly predicted as edge galaxies. We pick out these 12 images as shown in Fig.\ref{fig:fig_error_labeled}. It can be seen that the model has poor discrimination ability for samples with indistinct spiral arms due to high central brightness, and one galaxy will cause great interference to the identification of the other in a merge galaxy.
\begin{table}
	\centering
	\caption{ The classification accuracy when using $k$NN on the representation of the test sets.}
	\label{tab:acc_table}
	\begin{tabular}{lccccc} 
		\hline
		\multirow{2}{*}{Dataset} & \multicolumn{5}{c}{$k$ (\%) } \\
		& 1 & 5 & 10 & 15 & 20  \\
		\hline
		GZ2 &94.55 & 95.67 & 94.97 & 94.69 & 94.55\\
		SDSS-DR17 & 93.24 & 95.66 & 96.37 & 96.51 & 95.96 \\
		GZ DECaLS & 85.62 & 88.32 & 89.18 & 89.92 & 89.92 \\
		\hline
	\end{tabular}
\end{table}

\begin{figure}
	\includegraphics[width=1\linewidth]{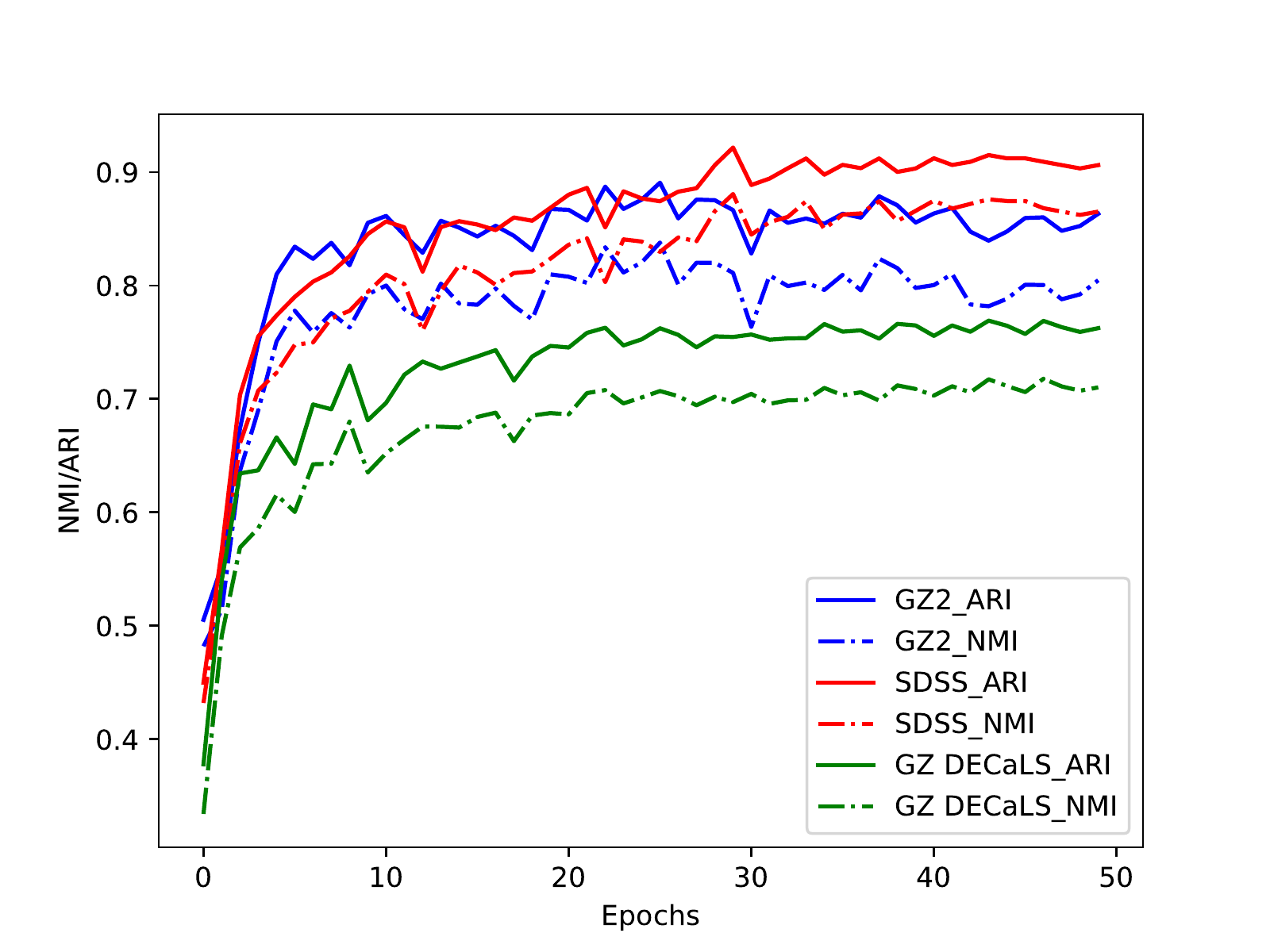}
    \caption{The trend diagram of ARI and NMI on validation sets as a function of the training epoch. The value gradually becomes stable as the training progresses, indicating that the clustering results also tend to be stable.}
    \label{fig:fig_ari_nmi}
\end{figure}

In order to measure the clustering performance, we compute ARI and NMI on validation sets after each epoch. The two value is almost close to 1 as shown in Fig.\ref{fig:fig_ari_nmi}. This proves that the predicted labels well match the real labels.

\subsection{Cross validation}
\begin{figure}
	\includegraphics[width=\columnwidth]{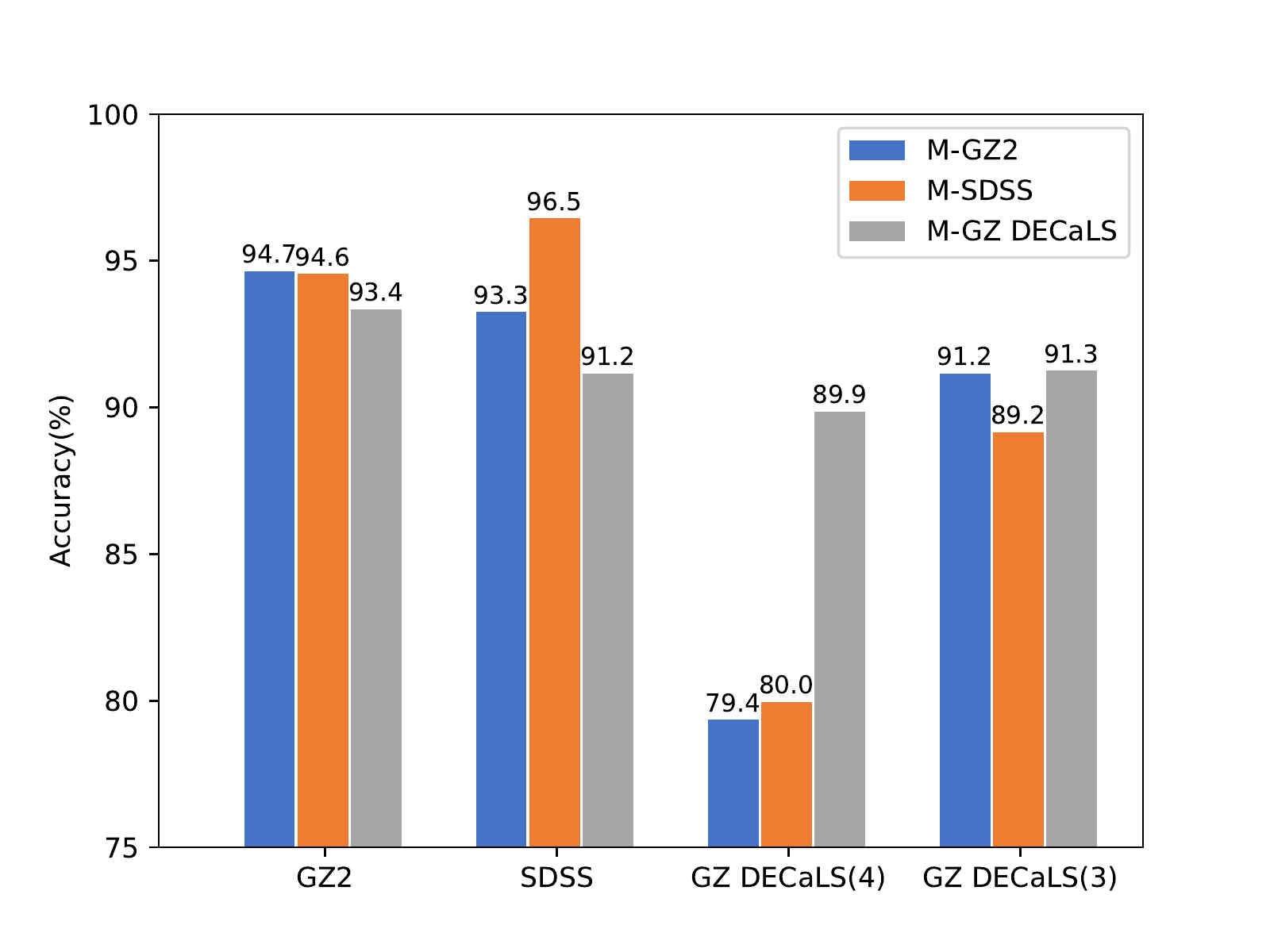}
    \caption{The accuracy of cross validation between GZ2, DECaLS and SDSS-DR17. The x-axis represents the test dataset used, DECaLS(4) with 4 categories (Round, Elliptical, Edge-on, and Spiral), and DECaLS(3) with 3 categories (Round, Edge-on, and Spiral). M-GZ2, M-SDSS and M-GZ DECaLS represents the model trained on the datasets of GZ2, SDSS, and GZ DECaLS respectively.}
    \label{fig:fig_cross_acc}
\end{figure}

In order to further validate the generalization ability of our model, we perform cross validation between the three dataset. That is applying the model trained upon one dataset to other two test dataset. The test accuracy values obtained only decline slightly as shown in Fig.\ref{fig:fig_cross_acc}.

Note that the accuracy of M-GZ2 and M-SDSS on DECaLS(4) achieve only 79.4\% and 80\%. This is because the test set of DECaLS(4) contains round classification, while the two models M-GZ2 and M-SDSS do not contain round classification during training. When DECaLS(3) excluded the round classification, and the accuracy increased to 91.2\% and 89.2\% respectively. For the decrease in number of classification leads to a reduction of the error rate, even with its own M-DECaLS, the accuracy also increased from 89.9\% to 91.3\%.

\subsection{Discussion of the result}
\subsubsection{Hierarchical clustering}
The visual difference of galaxies images is not obvious, which leads to the ambiguity of the boundary between the classification results. Thus, the accuracy based the labels from volunteers’ votes do not fully reflect the performance of the model. In order to get rid of the constraints of labels, we use the hierarchical clustering algorithm \citep{murtagh2012} to cluster the obtained sample feature representations. Then, we evaluate the clustering results to verify whether the model can truly cluster the most similar samples.

The process of hierarchical clustering is to initially treat each sample as a cluster (ie, leaf node). At each iteration, the two closest distances clusters polymerize to form a new cluster until one cluster is left. This idea coincides with the pairwise calculation of the cosine similarity between positive and negative samples in contrastive learning.
   
\begin{figure*}
	\includegraphics[width=\linewidth]{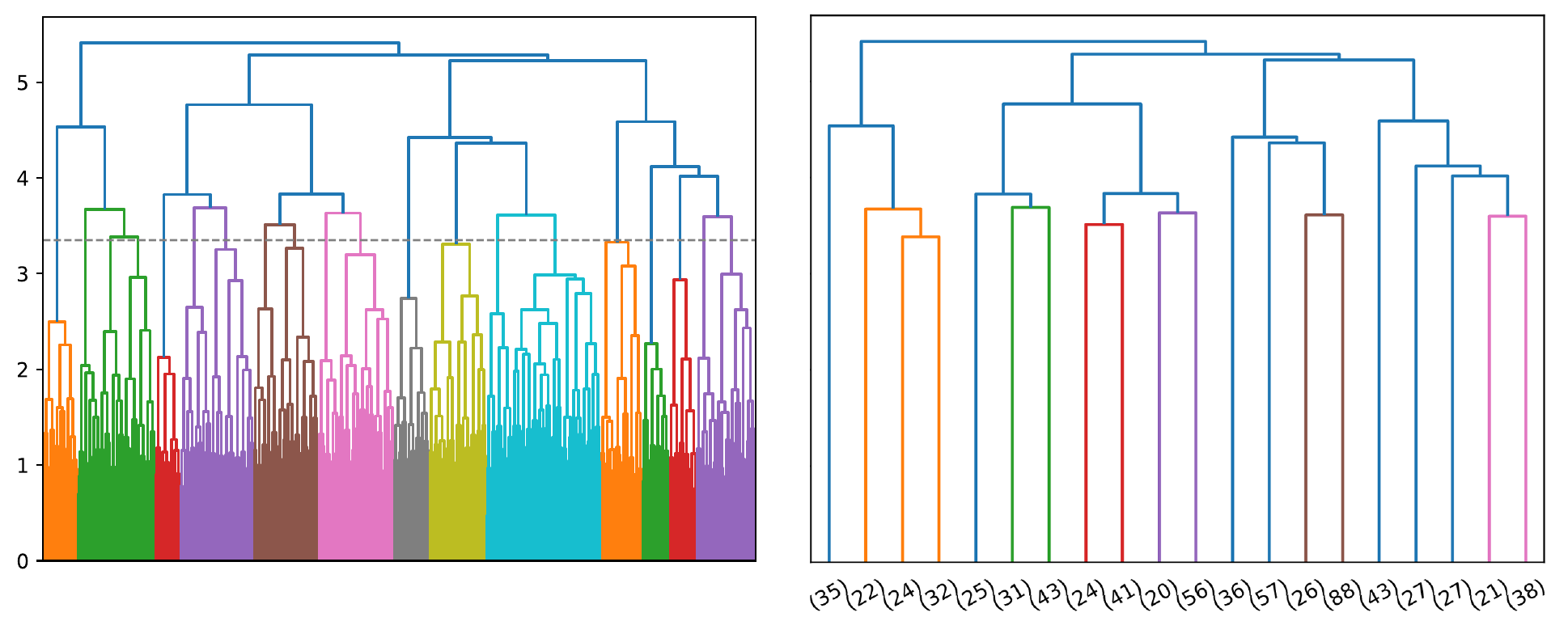}
    \caption{The results of hierarchical clustering on GZ2. The left panel represents the actual results of hierarchical clustering on all the test data from GZ2. We set the threshold of the distance between clusters to 3.3 as shown by the gray dotted line in the left panel, so as to obtain the schematic diagram of the right panel. The X-axis of the right panel indicates the number of samples of each cluster.}
    \label{fig:fig_hc_all}
\end{figure*}

Noted that in a hierarchical clustering tree, the closer to the root node, the greater the difference between cluster. Although the number of clusters is decreasing, it seems that this is for clustering, and it does not reflect the differences and similarities between each cluster. On the other hand, The closer to the leaf node, the more scattered the clusters are, which can not reflect the general characteristics of a certain classification. Therefore, we set an upper limit threshold of the distance between clusters, and select appropriate clustering results in the interlayer of the tree, for obtaining a more detailed and representative classification. Taking the test set of GZ2 as an example, we set this threshold to 3.3, and obtained 20 clusters (as shown in Fig.\ref{fig:fig_hc_all}).

\begin{figure*}
	\includegraphics[width=\linewidth]{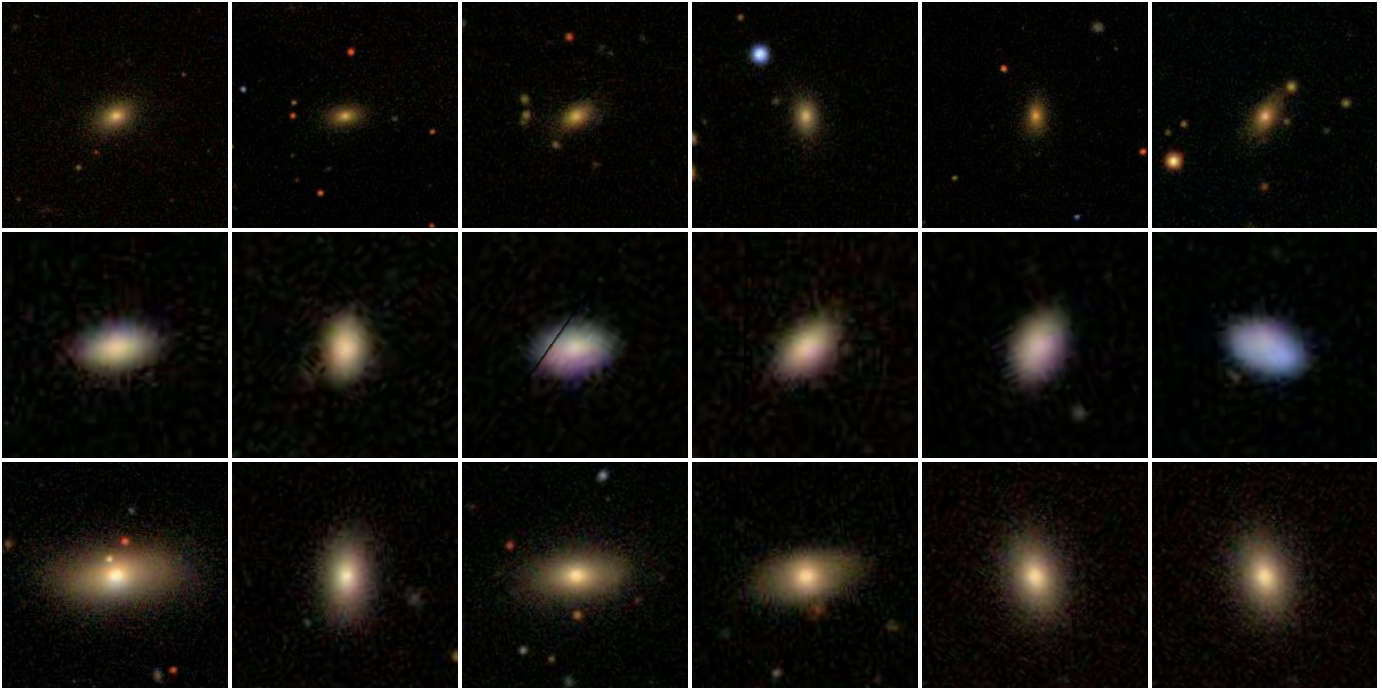}
    \caption{The selected 3 clusters after hierarchical clustering on the test dataset of GZ2. All samples of the 3 clusters are all elliptical galaxies.}
    \label{fig:fig_hierarchical}
\end{figure*}

We have selected 3 clusters which all are elliptical galaxies shown as Fig.\ref{fig:fig_hierarchical}, where the galaxies in the first row are smaller in shape, with a nucleus but less luminosity. The galaxies in the second row are slightly larger in shape but with blurred outlines, and the nuclei are less clear. The galaxies from the third row are the most luminous and with the most clear nuclei. That is, the elliptical galaxies can be divided into 3 clusters according to the morphological detailed features. It proves that our model has learned subtle morphological representations.

\subsubsection{Comparison with morphological measures}
To explore whether clustered galaxies also have intrinsic similarities on the quantified measures of galactic morphology, we make statistical studies on these measures after clustering. This test dataset was retrieved from the table \emph{PawlikMorph} in the SDSS skyserver at \url{http://skyserver.sdss.org/dr17/SearchTools/sql} (accessed May 7, 2022), which contains more than 4,200 galaxies and their measures, such as $A$, $G$ and $M_{20}$, etc. After filtered the null data, the used test set consists of 3228 galaxies. We have selected 6 common measures as described in Table \ref{tab:tab_pawlikmorph}. 
\begin{table*}
	\centering
	\caption{The 6 selected quantified measures of galactic morphology \citep{pawlik2014}.}
	\label{tab:tab_pawlikmorph}
	\begin{tabular}{ll} 
		\hline
		Measure & Description \\
		\hline
		$A$ & Asymmetry with the image rotating by $180^{\circ}$. \\
		$G$ & Gini index, is a measure of the inequality of the distribution of light within the galaxy. \\
		$M_{20}$ & Measuring the second-order moment of the brightest 20\% of the galaxy pixels.  \\
		$SB_0$ & S{\'e}rsic model best-fit parameter: the central surface brightness.  \\
		$R_{eff}$ & S{\'e}rsic model best-fit parameter: the effective radius.  \\
		$n$ & S{\'e}rsic model best-fit parameter: S{\'e}rsic Index.  \\				
		\hline
	\end{tabular}
\end{table*}

\begin{figure*}
	\includegraphics[width=\linewidth]{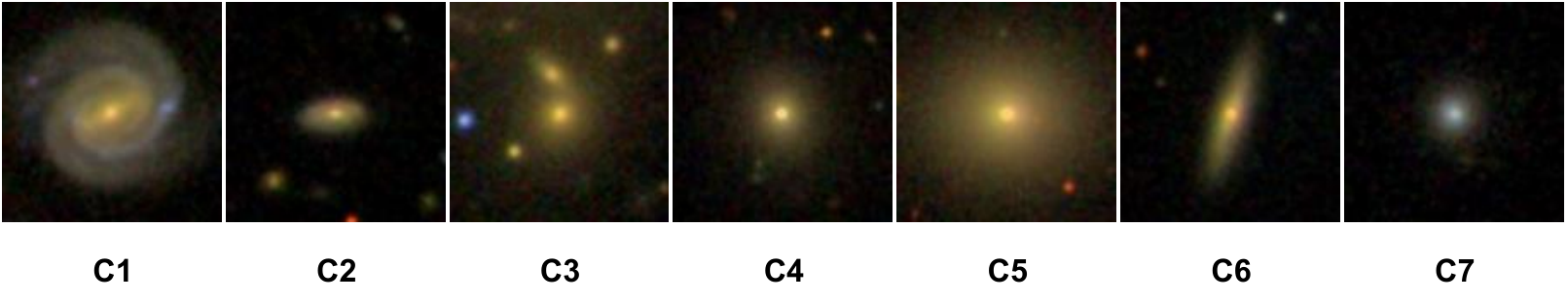}
    \caption{The representative sample from each cluster of 7 selected clusters. }
    \label{fig:fig_m_7}
\end{figure*}

These testing galactic images were fed into our trained model upon SDSS to obtain the feature representations on which hierarchical clustering was performed. With the condition of the distance threshold as 1, and we get 17 clusters from which we randomly selected 7 clusters for analysis. Table \ref{tab:tab_mean_pawlikmorph} lists the mean of 6 measures of galactic morphology on the 7 selected clusters, from which Fig.\ref{fig:fig_m_7} shows one representative image from each cluster. 
\begin{table*}
	\centering
	\caption{The mean of 6 measures of galactic morphology on 7 selected clusters.}
	\label{tab:tab_mean_pawlikmorph}
	\begin{tabular}{lllllllll} 
		\hline
		Cluster ID & $A$ & $G$ & $M_{20}$ &	$SB_0$ & $R_{eff}$ & $n$ & Size & Shape \\
		\hline
            C1 & 0.0782 & 0.5135 & -1.5042 & 14.8824 & 38.7451 & 1.9755 & 89 & Spiral \\
            C2 & 0.0671 & 0.6514 & -1.8845 & 29.5304 & 7.5086 & 2.1487 & 135 & Short-edge \\
            C3 & 0.1335 & 0.6367 & -1.5368 & 6.6808 & 27.0769 & 3.6885 & 148 & Merge \\
            C4 & 0.0637 & 0.6606 & -2.0805 & 15.6735 & 17.4898 & 3.2908 & 240 & Bright nucleus \\
            C5 & 0.0830 & 0.6620 & -1.7738 & 16.1455 & 28.0536 & 3.4286 & 71 & Bright nucleus /halo \\
            C6 & 0.0516 & 0.6377 & -2.0500 & 18.8785 & 12.7593 & 2.5856 & 131 & Long-edge \\
            C7 & 0.0623 & 0.6622 & -1.9596 & 24.7966 & 10.3644 & 2.7669 & 137 & Round/no nucleus \\
		\hline
	\end{tabular}
\end{table*}

\begin{figure*}
	\includegraphics[width=\linewidth]{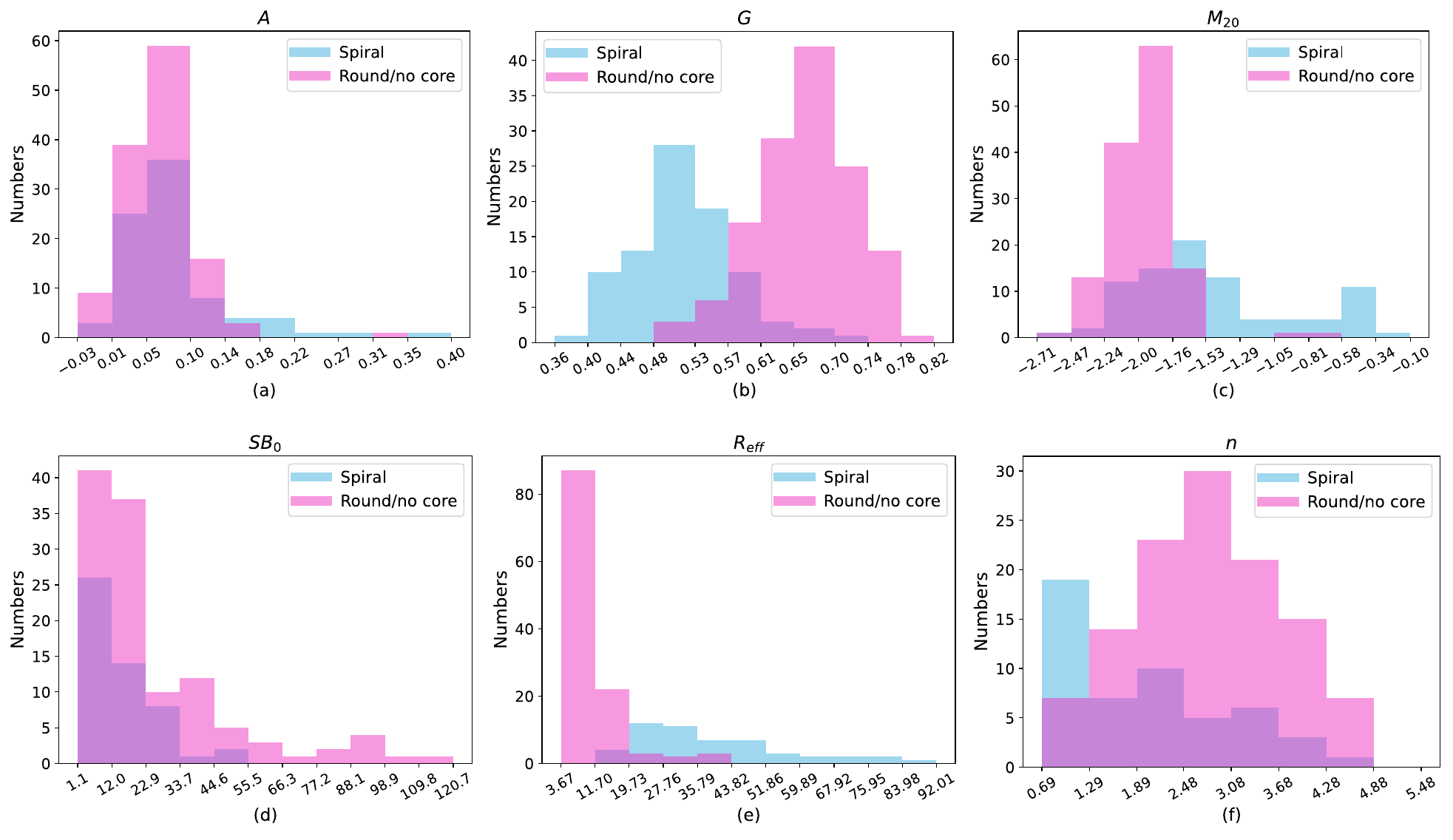}
    \caption{The comparison histogram of the 6 measures between C1 and C7, where the blue and pink histogram indicates C1 and C7 respectively.}
    \label{fig:fig_m_c1_c7}
\end{figure*}

Among 7 selected clusters, there is the biggest distance between C1 and C7. We compared the histogram of 6 measures between these two clusters in Fig.\ref{fig:fig_m_c1_c7}. The $M_{20}$ reflects the strength of luminosity convergence. The lower value represents the higher luminosity and the stronger convergence. Although the C1 has a slightly higher central luminosity, the convergence is reduced due to the larger shape. The overall luminosity of C7 is slightly weaker, the smaller shape increases the convergence. Therefore, so the $M_{20}$ distribution of C7 is generally leftward as shown in Fig.\ref{fig:fig_m_c1_c7}(c). In the same way, the C1 cannot concentrate the luminosity in the central area due to the existence of spiral arms. Hence the C1 is generally with small $G$ as shown in Fig.\ref{fig:fig_m_c1_c7}(b). The $R_{eff}$ reflects the size of shape. the $R_{eff}$ of C1 is larger than that of C7 as shown in Fig.\ref{fig:fig_m_c1_c7}(e). For the similarities of asymmetry, central surface brightness and S{\'e}rsic index between Spiral and Round, it also can been seen that there were no significant differences on $A$, $SB_0$ and $n$. From the above analysis, it is shown that our clustering results are well correlated with the structural measures of galaxies.

\subsubsection{Visualization Results}
\begin{figure*}
	\includegraphics[width=\linewidth]{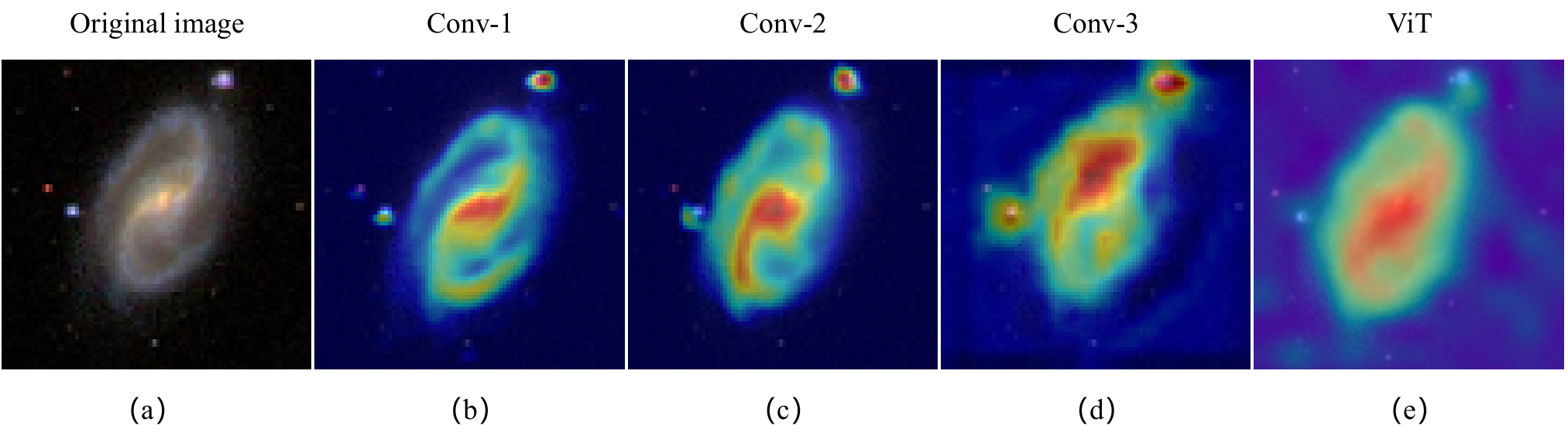}
    \caption{The original image and its generated feature maps through our model, where the sub-plots of Conv1-3 are generated from the 3 convolutional blocks of a Resnet-18 based network, and the sub-plot of ViT is  from the ViT block. }
    \label{fig:fig_heatmap}
\end{figure*}

Visualization of feature maps provide insight into the internal representation of a neural network. As discussed in Section \ref{sec:encoder}, our model adopted a fusing network with a Resnet-18 based convolutional network and a ViT block. For the convolutional network, we use Gradient-weighted Class Activation Mapping (Grad-CAM)\citep{Selvaraju2017} to extract the weight coefficients and gradients of each convolution layer to generate a heat map, which is then superimposed with the original image to generate a visual image. For the ViT block, we use the method proposed by \citet{Chefer202} to extract the attention map, then scale it to the original image size and overlay it with the original image to generate a visual image as shown in Fig.\ref{fig:fig_heatmap}.

The feature maps of first convolutional layer can response low-level patterns. As expected, Fig.\ref{fig:fig_heatmap}(b) shows obvious spiral contour. While the second and third convolutional layer provides high-level filters to show more detailed patterns as shown in Fig.\ref{fig:fig_heatmap}(c) and Fig.\ref{fig:fig_heatmap}(d). As can be seen from Fig.\ref{fig:fig_heatmap}(e) that the attention weights of the ViT block are mainly distributed on the spiral arms of the host spiral galaxy. This proves that our ViT block also pays attention to the backbone structure of the image.

\section{Conclusions}\label{sec:conclusions}

In this paper, we present an unsupervised method for learning galaxy visual representations and galaxy morphological classification. In order to learn contour information dominated images of galaxies, we design an encoder that fuse high-level and low-level features maps via ViT and CNN. We train the proposed model and evaluate the performance with sample images from GZ2, DECaLS and SDSS-DR17. The testing results of accuracy and clustering quality prove that our method has better performance. Even it can’t surpass supervised learning techniques for now, but yet it could be a powerful method that can be used in galaxy morphological classification where new surveys can produce massive unlabeled galactic images.

We performed cross validation to evaluate the robustness of the model. The results indicated the accuracy of cross validation only with a slight decrease. It demonstrates that transfer ability of our model from one survey to another. Analysis of the results through hierarchical clustering proves that our model can perform more detailed classification, which may also be helpful for correcting mislabeled images and discovering galaxies with new morphologies. We also study the distribution of typical structural measures of the clustering results with our method, and demonstrate the consistent correlation between the morphological classification results and the structural measures. The visualization of feature maps show that the \emph{Encoder} could capture the features of multiple-levels to provide rich semantic representation.

While the datasets used in this paper are mainly based on limited samples, the forthcoming surveys such as LSST and CSST will produce billions of galaxies which may include peculiar galaxies that cannot be associated with the known morphological classifications of the Hubble sequence. For future work, we would like to train this network on even larger datasets to investigate the efficiency of detecting peculiar galaxy. We also plan to take FITS as the input of training, since multiple band observational data intrinsically come with richer semantic information than RGB images. Although the current accuracy is above 90\%, it cannot compete with supervised learning. We expect the improvement of performance through using the state-of-the-art networks. We also plan to carry out a study on subtler classification of spiral type according to the tightness of spiral arms. 

\section*{Acknowledgements}
We thank the referee for useful comments that helped to improve the paper. This work was supported by the National Key Research and Development Program of China (2020SKA0110300), Funds for International Cooperation and Exchange of the National Natural Science Foundation of China (11961141001), and Joint Research Fund in Astronomy (U1831204, U1931141) under a cooperative agreement between the National Natural Science Foundation of China (NSFC) and Chinese Academy of Sciences (CAS), National Natural Science Foundation of China (No.11903009). The authors acknowledge financial support from the Yunnan Ten Thousand Talents Plan Young \& Elite Talents Project.

\bibliography{main}{}
\bibliographystyle{aasjournal}



\end{document}